# Survey of Latest Wireless Cellular Technologies for Enhancement of Spectral Density at Reduced Cost

Prof. R. K. Jain[1], Sumit katiyar[2] and Dr. N. K. Agrawal[3]

[1] Electronics Department, Singhania University, HIET Ghaziabad
Jhunjhunu, Rajasthan 333515, India

[2] Electronics Department, Singhania University, HIET Ghaziabad
Jhunjhunu, Rajasthan 333515, India

[3] EC Department, UPTU, IPEC
Ghaziabad, UP 201012, India

**Abstract**
The future of mobile wireless communication networks will include existing 3$^{rd}$ generation, 4$^{th}$ generation (implemented in Japan, USA, South Korea etc.), 5$^{th}$ generation (based on cognitive radio which implies the whole wireless world interconnection & WISDOM – Wireless innovative System for Dynamic Operating Megacommunications concept), 6$^{th}$ generation (with very high data rates Quality of Service (QoS) and service applications) and 7$^{th}$ generation (with space roaming). This paper is focused on the specifications of future generations and latest technologies to be used in future wireless mobile communication networks. However keeping in view the general poor masses of India, some of the future generation technologies will be embedded with 2G and 2.5G so that general masses may get the advantage of internet, multimedia services and the operators may get proper revenues with little extra expenditure in the existing mobile communication networks.

## 1. Introduction

Mobile devices together with the intelligence that will be embedded in human environments - home, office, public places - will create a new platform that enables ubiquitous sensing, computing, storage and communication. Core requirements for this kind of ubiquitous ambient intelligence are that the devices are autonomous and robust. They can be deployed easily and require little maintenance. As shown in Fig 1, mobile device will be the gateways to personally access ambient intelligence and needed information. Mobile also implies limited size and restrictions on the power consumption. Seamless connectivity with other devices and fixed networks is a crucial enabler for ambient intelligence system - this leads to requirements for increased data-rates of the wireless links.

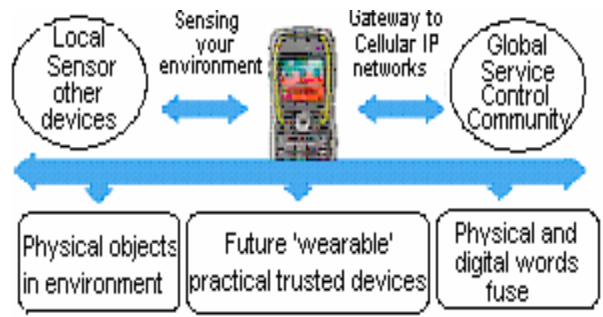

Fig.1 Mobile devices become gateway to ambient intelligence and needed information

Intelligence, sensing, context awareness and increased data-rates require more memory and computing power, which together with the size limitations leads to severe challenges in thermal management. All the above requirements can be addressed satisfactorily with the application of OFDM, CDMA-2000, WCDMA/UMTS, TD-SCDMA, Wi-Fi (i.e. Wireless LAN) networks with fixed internet to support wireless mobile internet as the same quality of service as fixed internet, which is an evolution not only to move beyond the limitations and problems of 3G, but also to enhance the quality of services, to increase the bandwidth and to reduce the cost of the resource, 5G based on cognitive radio, 6G (to integrate satellites for getting global coverage) and nanotechnology [1].

4G mobile systems will mainly be characterized by a horizontal communication model, where such different access technologies as cellular, cordless, wireless LAN type systems, short-range wireless connectivity, and wired systems will be combined on a common platform to complement each other in the best possible way for different service requirements and radio environments [2].

The 5th wireless mobile multimedia internet networks can be completed wireless communication without limitation, which bring us perfect real world wireless – World Wide Wireless Web (WWWW). 5G is based on 4G technologies, which is to be revolution to 5G. During this processing, there are two kind of problems need to be solved. The first is wider coverage and the second is freedom of movement from one technology to another. The 6$^{th}$ generation (6G) wireless mobile communication networks shall integrate satellites to get global coverage. The global coverage systems have been developed by four courtiers. The global position system (GPS) is developed by USA. The COMPASS system is developed by China. The Galileo system is developed by EU, and the GLONASS system is developed by Russia [3]. These independent systems are difficulty for space roaming. The task of 7th generation (7G) wireless mobile communication networks are going to unite the four systems to get space roaming.

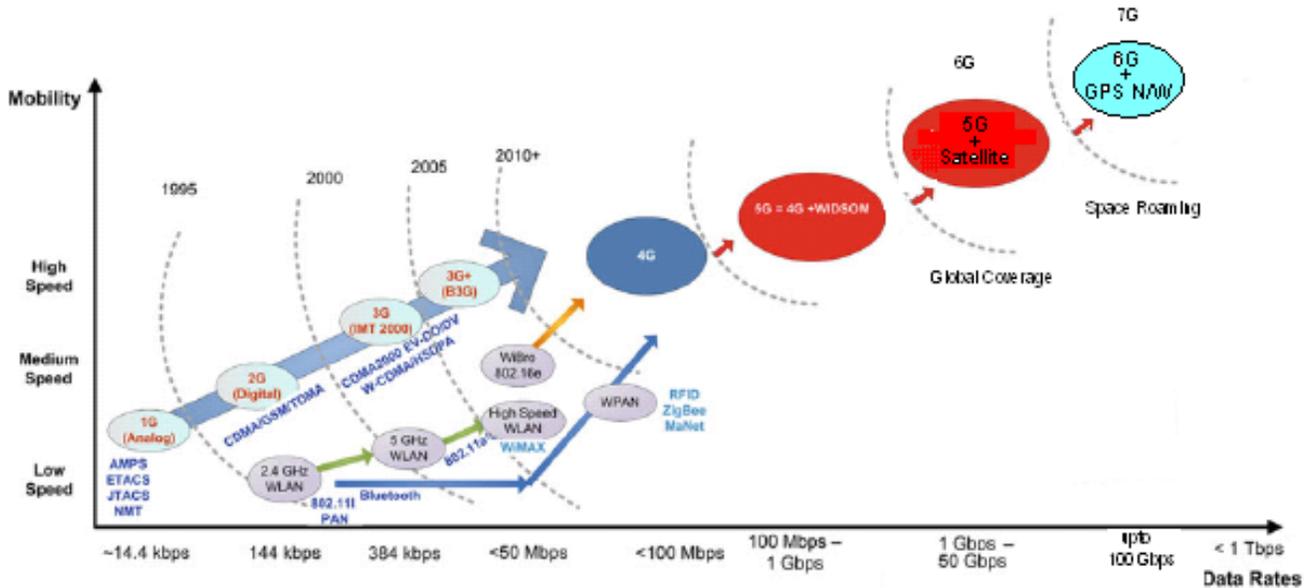

Fig.2 Evolution towards 7G

## 2. Review

### 2.1 First Generation (1G)-Analog System

The first generation wireless mobile communication system is not digital technology, but analog cellular telephone system which was used for voice service only during the early 1980s [4]. This Advanced Mobile Phone System (AMPS) was a frequency modulated analog mobile radio system using Frequency Division Multiple Access (FDMA) with 30kHz channels occupying the 824MHz − 894MHz frequency band and a first commercial cellular system deployed until the early 1990's [5].

### 2.2 Second Generation (2G)-Digital System

|  | *AMPS* | *D-AMPS* | *GSM* | *CDMA* |
|---|---|---|---|---|
| Operating Spectrum Frequency | 800 MHz | 800 & 1900 MHz | 900 & 1800 MHz (Europe) 800 & 1900 MHz (US) | 800 & 1900 MHz |
| Channel | 30 | 30 | 200 | 1.25 |
| Width | KHz | KHz | KHz | MHz |
| User Per Channel | 1 | 3 | 8 | About 20 |
| Channel Separation | Frequency | Frequency And Time | Frequency And Time | Frequency and Code |
| Network Architecture | IS-41 | IS-41 | GSM-MAP | IS-41 |

### 2.3 Comparison between 2.5G and 3G

| *Items* | *2.5G* | *3G* |
|---|---|---|
| Speed | Up to 384 Kbps | Up to 2Mbps |
| Databases | HLR, VLR, EIR, AuC | Enhanced HLR, VLR, EIR, AuC |
| Core Network | Circuit and packet switching | Wide-area concept Circuit and packet switching |
| Technologies | HSCSD GPRS EDGE | WCDM, CDMA2000, TD-SCDMA |
| Applications | SMS, Internet | Internet, multimedia |

## 2.4 Comparison between 4G and 5G

| Items | 4G | 5G |
|---|---|---|
| Speed | Up to 1 Gbps | Up to 1 Gbps on mobile |
| Services | Global Roaming | Global Roaming Smoothly |
| Core Network | Broadband, Entirely IP-based packet switching | Enhanced Broadband Entirely IP-based packet switching |
| Technologies | OFDM, MC-CDMA, LAS-CDMA, Network-LMPS | LAS-CDMA, OFDM, MC-CDMA, UWB, IPv6 Network-LMDS |

# 3. Fourth Generation (4G): Revolutionary Approach

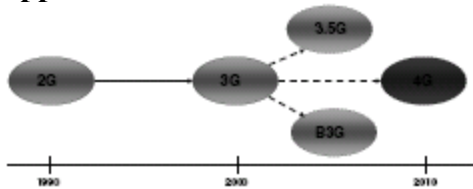

- Higher bit rates than 3G and full mobility.
- Higher spectral efficiency and lower cost per bit than 3G
- Air Interface optimized for IP traffic.
  Examples: OFDM, MIMO.

### 3.1 Technical Issues

1. High data rates- OFDM, MC-CDMA, Synchronization & estimation, Distortion (linear, non-linear).
2. Coding- Iterative decodable codes (Turbo, LDPC)
3. Smart antenna systems
4. MIMO (Multi Input Multi Output) devices
5. Reconfigurable terminals (SW and HW)
6. Cognitive Radio

### 3.2 Reconfigurable Technology

Reconfigurable refers to the software re-definition and/or adaptation of every element within each layer of the communication chain.

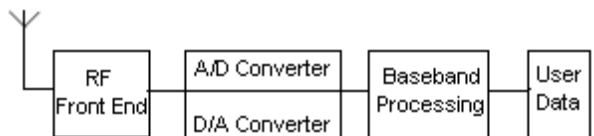

Fig .3 Software Driven Radio

### 3.3 OFDM

3G systems such as high speed packet access (HSPA) provide up to round 15-20 Mbps downlink and about 5-10 Mbps uplink. 4G systems are being designed to support 5 to 10 times these rates (i.e. downlink above 100 Mbps and uplink above 50 Mbps). OFDMA is based on orthogonal frequency division multiplexing. At first OFDM was used into fixed access WIMAX 802.16D to provide high speed internet access either as a replacement for other access technologies like ADSL or cable, or to provide service in regions where the other access technologies where not deployed.

The widespread interests of OFDM become clear from a glance at OFDM characteristics. In 802.11a, OFDM provides raw data rates up to 54 Mbits / s in a 20-MHz channel. In addition to supporting high data capacity and resisting degradation from various types of radio effects, OFDM makes highly efficient use of the available spectrum. The latter characteristic will become crucial in coming years as wireless networks are built out.

OFDM Simple Architecture Overcomes - Noise, Signal to Noise Ratio Challenges, Multipath Fading, Adjacent Channel, Interference, Non-Adjacent Channel Interference OFDM also provides a frequency diversity gain, improving the physical layer performance. It is also compatible with other enhancement technologies such as smart antennas and MIMO. OFDM modulation can also be employed as a multiple access technology (OFDMA). In this case, each OFDM symbol can transmit information to/from several users using a different set of sub carriers (sub channels). This not only provides additional flexibility for resource allocation (increasing the capacity), but also enables cross-layer optimization of radio link usage.

The idea of the complementation of IPv6, OFDM, MC-CDMA, LAS-CDMA, UWB and Network-LMDS can be arranged in different zone size. IPv6 can be designed for running in the all area because it is basic protocol for address issue. LAS-CDMA can be designed for the global area as zone 1, world cell. OFDM and MC-CDMA can be designed for running in the wide area (Zone 3), called Macro cell. Network-LMDS is in Zone 2, Micro cell, and UWB is in Zone 1, Pico cell.

MC-CDMA stands for Multi-Carrier Code Division Multiple Access, which is actually OFDM with a CDMA overlay LAS-CDMA Large Area Synchronized Code Division Multiple Access is developed by Link Air Communication, a patented 4G wireless technology. "LAS-CDMA enables high-speed data and increases voice capacity and the latest innovative solution.

In 4G technologies, UWB *7 radio can help solve the multi-path fading issues by using very short electrical pulses to across all frequencies at once.

The Network-LMDS, Local Multipoint distribution system, is the broadband wireless technology used to carry voice, data, Internet and video services in 25GHz and higher spectrum.

### 3.4 Multiple-Input Multiple-Output

MIMO uses signal multiplexing between multiple transmitting antennas (space multiplex) and time or frequency. It is well suited to OFDM, as it is possible to process independent time symbols as soon as the OFDM waveform is correctly designed for the channel. This aspect of OFDM greatly simplifies processing. The signal transmitted by m antennas is received by n antennas. Processing of the received signals may deliver several performance improvements: range, quality of received signal and spectrum efficiency.

### 3.5 Software Defined Radio

Software Defined Radio (SDR) benefits from today's high processing power to develop multi-band, multi-standard base stations and terminals. Although in future the terminals will adapt the air interface to the available radio access technology, at present this is done by the infrastructure. Several infrastructure gains are expected from SDR. For example, to increase network capacity at a specific time (e.g. during a sports event), an operator will reconfigure its network adding several modems at a given Base Transceiver Station (BTS).

SDR makes this reconfiguration easy. In the context of 4G systems, SDR will become an enabler for the aggregation of multi-standard pico/micro cells. For a manufacturer, this can be a powerful aid to providing multi-standard, multi-band equipment with reduced development effort and costs through simultaneous multi-channel processing.

## 4. 5G Based on Cognitive Radio

### 4.1 5G Concept

The twenty-first century is surely the "century of speed", and achieves a high evolution in all the possible domains, especially in communication: a very large variety of services, software, equipments, possibilities etc. But this huge and colored offer also brings a complicated lifestyle and waste of time for the human beings, and needs to be integrated and achievable in a simple manner. Therefore, a new technology started to be delineated, that will provide all the possible applications, by using only one universal device, and interconnecting the already existing communication infrastructures—that is the fifth generation of the mobile communications standards—5G.

Both the cognitive radio (CR) and the fifth generation of cellular wireless standards (5G) are considered to be the future technologies: on one hand, CR offers the possibility to significantly increase the spectrum efficiency, by smart secondary users (CR users) using the free licensed users spectrum holes; on the other hand, the 5G implies the whole wireless world interconnection (WISDOM—Wireless Innovative System for Dynamic Operating Mega communications concept), together with very high data rates Quality of Service (QoS) service applications.

Cognitive Radios (CRs) integrate radio technology and networking technology to provide efficient use of radio spectrum, a natural resource, and advanced user services.

The idea of a cognitive radio extends the concepts of a hardware radio and a software defined radio (SDR) from a simple, single function device to a radio that senses and reacts to its operating environment.

A Cognitive Radio incorporates multiple sources of information, determines its current operating settings, and collaborates with other cognitive radios in a wireless network. The promise of cognitive radios is improved use of spectrum resources, reduced engineering and planning time, and adaptation to current operating conditions. Some features of cognitive radios include:

*Sensing the current radio frequency spectrum environment:* This includes measuring which frequencies are being used, when they are used, estimating the location of transmitters and receivers, and determining signal modulation. Results from sensing the environment would be used to determine radio settings.

*Policy and configuration databases*: Policies specifying how the radio can be operated and physical limitations of radio operation can be included in the radio or accessed over the network. Policies might specify which frequencies can be used in which locations. Configuration databases would describe the operating characteristics of the physical radio. These databases would normally be used to constrain the operation of the radio to stay within regulatory or physical limits.

*Self-configuration*: Radios may be assembled from several modules. For example, a radio frequency front-end, a digital signal processor, and a control processor. Each module should be self-describing and the radio should automatically configure itself for operation from the available modules. Some might call this "plug-and-play."

*Mission-oriented configuration*: Software defined radios can meet a wide set of operational requirements. Configuring a SDR to meet a given set of mission requirements is called mission oriented configuration. Typical mission requirements might include operation within buildings, substantial capacity, operation over long distances, and operation while moving at high speed. Mission-oriented configuration involves selecting a set of radio software modules from a library of modules and connecting them into an operational radio.

*Adaptive algorithms*: During radio operation, the cognitive radio is sensing its environment, adhering to policy and configuration constraints, and negotiating with peers to best utilize the radio spectrum and meet user demands.

*Distributed collaboration*: Cognitive radios will exchange current information on their local environment, user demand, and radio performance between themselves on regular bases. Radios will use their local information and peer information to determine their operating settings.

*Security*: Radios will join and leave wireless networks.

## 5. Benefit of Nanotechnology

Mobility also implies limited size and restriction on the power consumption. Seamless connectivity with other devices and fixed networks is a crucial enabler for ambient intelligence systems- this leads to requirements for increased data rates of the wireless links. Intelligence, sensing, context awareness, and increased data rates require more memory and computing power, which together with the size limitations leads to severe challenges in thermal management.[10], [11]
All these requirements combined lead to a situation which can not be resolved with current technologies. Nanotechnology could provide solutions for sensing, actuation, radio, embedding intelligence into the environment, power efficient computing, memory, energy sources, human-machine interaction, materials, mechanics, manufacturing, and environmental issues [6].

## 6. Hierarchical System

The vision of the "third generation" cellular system incorporates micro & picocells for pedestrians use, with macro cells for roaming mobiles. In order to increase the growing capacity demands of cellular mobile communication systems cell splitting will be applied and/or small pico cell will be established .Since both measures can increase spectral efficiency. Hierarchical cellular networks have been suggested previously to overcome the inherent disadvantage of an increased no of handoffs, which both cell splitting and small pico cells, bring about. A critical question with respect to hierarchical cellular networks is how to divide the available radio resources (i.e. frequency, channels) among the macro and micro cells layers in a optimal way. Another important aspect is the optimal choice of a threshold velocity above which users are assigned to the macro cell layer. Most research in this area so far has dealt with those issues is a static way, assuming fixed traffic and mobility parameters. First time in the year 2000, two adaptive algorithms are described, which control the threshold velocity as well as the division of the resources among these layers, dynamically. The performance of those algorithms is evaluated by means of computer simulations [7].

### 6.1 Macro Cell

A conventional base station with 20W power and range is about 20 km to 30 km.

### 6.2 Micro Cell

A conventional base station with 5W power and range is about 1km to 5 km.

### 6.3 Pico Cell

The Pico cells are small versions of base stations, ranging in size from a laptop computer to a suitcase. Besides plugging coverage holes, Pico cells are frequently used to add voice and data capacity, something that repeater and distributed antenna can not do.

Adding capacity in dense area, splitting cells are expensive, time consuming and occasionally impossible in dense urban environment where room for a full size base station often is expensive or unviable. Compact size Pico cells makes them a good fit for the places needing enhanced capacity, they can get.

Picocells are designed to serve very small area such as part of a building, a street corner, malls, railway station etc. These are used to extend coverage to indoor area where outdoor signals do not reach well or to add network capacity in areas with very dense uses.

## 7. Proposed Network

Based on smart antenna, MC-CDMA and OFDMA technologies cellular network has been proposed with minor amendments in existing 2G and 2.5G systems for providing quality service and enhanced spectral density at reduced rate.[9], [12], [13], [14], [15], [16] [17].

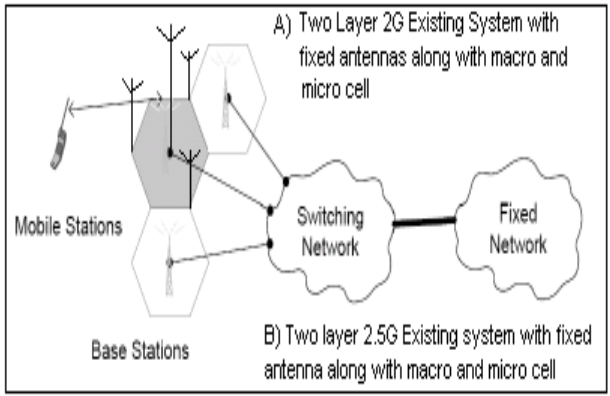

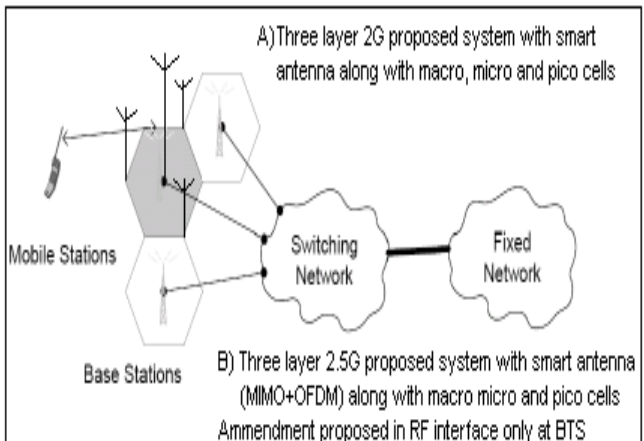

Fig.4 Proposed Network

## 8. Conclusion

In this paper the survey of 1G to 4G, 5G and CR technologies have been presented & discussed. The important technologies required for achieving desired specifications were also discussed and finally roadmap for probable 5G, 6G and 7G networks have been proposed.
5G obtains the tool technology to interconnect and integrate all the wireless network types and thus gains the needed network infrastructure (CR network).

6G will integrate all wireless mobile networks with satellites to get global coverage.

7G wireless mobile networks are going to unite the four GPS systems available in space to get space roaming in addition to 6G systems.

However our complete attention is to enhance the quality of services, to increase the bandwidth and to reduce the cost of the resource [8] along with reduction of RF pollution and power consumption.

The modified networks for 2G and 2.5G systems have been proposed to get the desired results. The addition of smart antenna in the system alone can increase spectral efficiency and quality of services manifold.

Prof. R. K. Jain has received his BE and ME in 1972 & 1974 from IIT Roorkie, respectively. He has 32 years industrial experience from M/S ITI Ltd. (Top Management) and 5 years teaching experience as professor. He is currently working as Prof. and Head of Electronics & Communication Engineering at HIET, Ghaziabad, UP, India. He is currently pursuing PhD degree course from Singhania University as research scholar. His area of research is cellular communication. He has published 6 papers in National / International Conferences.

Sumit Katiyar has received his B.Sc. and M.Sc. in 2001 & 2004 respectively from CSJMU, Kanpur, UP, India. He has received his M.Tech. from JIITU, Noida, UP, India. . He is currently pursuing PhD degree course from Singhania University as research scholar. He is currently working as Sr. Lecturer in department of Electronics & Communication Engineering at HIET, Ghaziabad, UP, India.

Dr. N. K. Agrawal is a senior member of IEEE (USA) and life member of ISTE and ISCEE.